\documentclass[12pt]{article}
\begin{document}

\begin{titlepage}
\begin{center}
{\Large \bf Interplay between the small and the large scale structure of spacetime }

\vspace{5mm}

\end{center}

\vspace{5 mm}

\begin{center}
{\bf Moninder Singh Modgil
\footnote[1]{PhD,  in Physics, from Indian Institute of Technology, Kanpur, India, and
B.Tech. (Hons.) in Aeronautical Engineering, from Indian Institute of Technology, Kharagpur, India.
\\ email: msmodgil@gmail.com}}
  
\vspace{3mm}

\end{center}

\vspace{1cm}

\begin{center}
{\bf Abstract}
\end{center}
Existence of  frame invariant, maximum, time interval $T$, length $L$, and mass $M$ is postulated. In the de Sitter universe - (1) the life span of universe, (2) the circumference of universe at the point of maximum expansion, and (3) the mass of the universe - are candidates for $T$, $L$ and $M$ respectively. Impact of such invariant global parameters, on the definition of local physical quantities, such as velocity is discussed.

\vspace{1cm}

\end{titlepage}

\section{Introduction: the maxim of maximum, invariant - time, length and mass}

Segal \cite{Segal 1976} noted that a possible route to investigations of nature, is to assign a finite magnitude to a quantity, which tacitly is assigned an infinite value. This was motivated by the observation that the maximum possible speed - namely that of light - is finite (rather than infinite) and invariant in all reference frames. A variation on this theme, is ascribing non-zero values to certain quantities, such as the Plank's constant in quantum mechanics. A zero Plank's constant takes away the concept of uncertainty principles and the noncommuting variables, of the quantum mechanics, and relegates it to classical mechanics. Other examples are discrete spacetimes, in which the infinitesmal space and time intervals, are replaced by finite intervals -  Snyder spacetime \cite{Snyder 1947} being one such case.  

The Newtonian universe has $R^3$ spatial topology, and $R^1$ temporal topology. Both the space and time coordinates,  range in the interval $(-\infty, \infty)$. There is no maximum  space or time interval - and  no maximum speed, in the Newtonian universe. The Minkowski spacetime with $R^3 \times R^1$ topology and indefinite metric, had a maximum invariant speed - namely that of light, but no maximal spatial or temporal interval. The Einstein universe has $S^3$ spatial topology and $R^1$ temporal topology. Thus, while the maximum spatial interval was restricted by the size of the universe there however, was no maximum time interval, in the Einstein universe.  Segal suggested $S^3 \times S^1$ topology for spacetime, i.e., Einstein's universe with a finite periodic time. The cosmological red shift in Segal's model is obtained via the \textit{uni-energy} operator. In the context of periodic time, we note parenthetically that, it is fairly routine to have an imaginary periodic time, in  finite temperature field theory \cite{Rey 2006, Kleeban 2004} and Euclidean quantum gravity \cite{Kim 1999, Smolin 1995}.  

Time, length and mass are among the fundamental physical dimensions. Physicists routinely employ dimensional check for terms in an equation, to ensure their validity. Could it be that there exist, fundamental, maximum, and frame invariant - time interval $T$, spatial length $L$, and  mass $M$? In the de Sitter universe,  it does  makes sense to talk about - (1) a maximum physical time interval $T$ equal to the life span\footnote{The life span of the universe is the period between the big bang, and the big crunch. It is different from the age of universe, measured at a specific time - except the age at the time of the demise of the universe.} of the universe, (2) a maximum spatial interval or length $L$ equal to the circumference of universe $L$ at the point of maximum expansion, and (3) a maxium mass $M$ equal to the mass of the universe. They could also be suitably defined in other cosmological models. Consider the maxim that - \textit{The global parameters $T$, $L$ and $M$, of structure of spacetime, are frame invariant.} It is shown in this paper that, this maxim requres a new definition of velocity $v_{TLM}$, depending upon these global parameters.

\section{Interplay between the local and the global}

From an operational view point, velocity is measured by noting location of an object at the spatial point $s_1$ on time instant $t_1$, - and then at a spatial point $s_2$, at time instant $t_2$. This gives the physically measured velocity $v$ -

\begin{eqnarray} \label{Velocity operational definition}
v=\frac{s_2 - s_1}{t_2 - t_1} = \frac{\Delta s}{\Delta t}
\end{eqnarray}

\noindent \noindent $\Delta s$ and $\Delta t$ are subject to measurement uncertainties of classical and quantum mechanical nature. Newton essentially initiated the invention of calculus in the process of defining velocity $v_{Newton}$ as the ratio of vanishingly small space interval $\Delta s$, and time interval $\Delta t$ -

\begin{eqnarray}\label{Velocity definition}
	v_{Newton}=\frac{\lim_{\Delta s \rightarrow 0} \Delta s}{\lim_{\Delta t \rightarrow 0} \Delta t} = \frac{ds}{dt}
\end{eqnarray}

\noindent This definition is local and independent of any large scale, global parameters of the structure of spacetime. Now, the Lorentz transformation linking a frame $\mathcal{F'}$, moving with relative velocity $v$ with respect to a frame $\mathcal{F}$, actually depend upon the velocity defined in equation [\ref{Velocity definition}].  Time intervals, spatial lengths, and mass - measured in the two frames are related by the well known expression -

\begin{eqnarray}\label{Lorentz transformation Time}
	\Delta t' = \frac{\Delta t}{\sqrt{1-\frac{v^2}{c^2}}}
\end{eqnarray}

\begin{eqnarray}\label{Lorentz transformation Length}
	\Delta s' = \frac{\Delta s}{\sqrt{1-\frac{v^2}{c^2}}}
\end{eqnarray}

\begin{eqnarray}\label{Lorentz transformation Mass}
	m' = \frac{m}{\sqrt{1-\frac{v^2}{c^2}}}
\end{eqnarray}

\noindent Here, superscript $'$ refers to quantities defined in $\mathcal F'$, while those without this superscript refer to quantities defined in $\mathcal{F}$.  Invariance of $T$, $L$, and $M$ between $\mathcal{F}$ and $\mathcal{F'}$ is given by the conditions -

\begin{eqnarray}
	T=(\Delta t')_{max} = (\Delta t)_{max}
\end{eqnarray}

\begin{eqnarray}
	L=(\Delta s')_{max} = (\Delta s)_{max}
\end{eqnarray}

\begin{eqnarray}
	M=m'_{max} =  m_{max}
\end{eqnarray}

\noindent These conditions demand a modification of the factor $\gamma_{STR}$ of Special Theory of Relativity (STR) -

\begin{eqnarray}
\gamma_{STR} =\frac{ 1}{\sqrt{1-v^2_{Newton}/c^2}}.
\end{eqnarray}

\noindent as otherwise there would exists - time intervals longer than $T$, spatial intervals larger than $L$, and masses larger than $M$, i.e., -

\begin{eqnarray}
 \Delta t' = \gamma_{STR} (\Delta t = T)   > T
\end{eqnarray}

\begin{eqnarray}
  \Delta s' = \gamma_{STR} (\Delta s = L)  > L
\end{eqnarray}

\begin{eqnarray}
 m' = \gamma_{STR} (m = M) > M
\end{eqnarray}

Lets first consider the largest physical time interval measurable in the de Sitter universe - namely its life span. Consider a set of clocks which start ticking at big bang, and go on ticking till the big crunch. No matter how the clocks move, i.e., with different velocities, our maxim requires that, they all show the same elapsed time $T$, between the big bang and the big crunch. What we are posing is - \textit{a cosmological version of twins paradox, in which the time interval of travel equals the life span of the universe}\footnote{In this context, it would be relevant to mention work of Boblest, M\"{u}ller and Wunner \cite{Boblest 2010}, on twin paradox in de Sitter universe, and that of Barrow and Levin \cite{Barrow 2001} on twin paradox in compact spaces.}.   Evidently, this requires that, the constant $T$, the measured time variable $\Delta t$ and $\Delta s$, appear in any equation which relates time intervals between different reference frames. As the velocity $v$ is not time dependent (i.e., we are not considering frames which are accelerating with respect to each other), equation [\ref{Lorentz transformation Time}] can be re-written as -

\begin{eqnarray}
	\Delta t' = \frac{\Delta t}{\sqrt{1-\frac{1}{c^2} \left( \frac{\Delta s}{ \Delta t} \right)^2}}
\end{eqnarray}

\noindent where, $\Delta s = v \Delta t$ is the distance travelled in the time interval $\Delta t$. One way to obtain this frame independence of $T$, is by modifying equation [\ref{Lorentz transformation Time}] to -

\begin{eqnarray}
	\Delta t' = \frac{\Delta t}{\sqrt{1-\frac{1}{c^2} \left[ \left( \frac{\Delta s}{ \Delta t} \right) \left(1- \frac{\Delta t}{ T} \right) \right]^2}}
\end{eqnarray}

\noindent To retain the form of equation [\ref{Lorentz transformation Time}], following definition for the velocity $v_T$ (for an invariant maximum time interval $T$), can be adopted -

\begin{eqnarray}\label{Velocity definition T}
 v_T = v_{Newton} \left( 1 - \frac{\Delta t}{T} \right)
\end{eqnarray}

\noindent Note that, 

\begin{eqnarray}\label{Velocity Newtonian limit}
	v_{Newton} = \lim_{T \rightarrow \infty} v_T (\Delta s, \Delta t) 
\end{eqnarray}

\noindent and,

\begin{eqnarray}\label{Recurrence constraint}
	v_T(\Delta s, \Delta t = T) = 0
\end{eqnarray}

\noindent  Equation [\ref{Recurrence constraint}] essentially implies that the change in position of a particle, over a period $T$ (recurrence period) is zero, i.e., the position of object in snap shots seperated by the time interval $T$, appears unchanged -  alternatively, the object appears stationary. This, thus satisfies the criterion for recurrence in a periodic universe, indicated in \cite{Modgil 2009}. Clearly for non-vanishing $\Delta s$, and finite $T$, 

\begin{eqnarray}\label{Velocity comparision}
	v_T (\Delta s, \Delta t) < v_{Newton}
\end{eqnarray}

\noindent Note also that, $\Delta t$ is greater than $T$ for superlumical velocities $v (>c)$, 

\begin{eqnarray}
 \Delta t' > T \Rightarrow v > c \sqrt{\frac{T+\Delta t}{T-\Delta t}}
\end{eqnarray}

Proceeding as in case of $T$ above, frame invariance of $L$ and $M$, can be achieved by defining velocity as -

\begin{eqnarray}\label{Velocity definition L}
v_L(\Delta s, \Delta t) = v_{Newton} \left( 1 - \frac{\Delta s}{L} \right) 
\end{eqnarray}

\begin{eqnarray}\label{Velocity definition M}
v_M (\Delta s, \Delta t, m)= v_{Newton} \left( 1- \frac{m}{M} \right) 
\end{eqnarray}

\noindent Equations [\ref{Velocity definition T}], [\ref{Velocity definition L}] and [\ref{Velocity definition M}] can be combined to arrive at the following definition of velocity $v_{TLM}$ - 

\begin{eqnarray}\label{Velocity definition TLM}
v_{TLM}(\Delta s, \Delta t, m) = v_{Newton} \left( 1 - \frac{\Delta t}{T} \right) \left( 1 - \frac{\Delta s}{L} \right) \left( 1 - \frac{m}{M} \right)
\end{eqnarray}

\noindent The corresponding special relativistic factors for transforming between the frames $\mathcal{F}$ and $\mathcal{F'}$ are -

\begin{eqnarray}\label{gamma T}
\gamma_T = \frac{1}{\sqrt{1-\frac{1}{c^2} \left[ \left( \frac{\Delta s}{ \Delta t} \right) \left(1- \frac{\Delta t}{T} \right) \right]^2}}
\end{eqnarray}

\begin{eqnarray}\label{gamma L}
\gamma_L = \frac{1}{\sqrt{1-\frac{1}{c^2} \left[ \left( \frac{\Delta s}{ \Delta t} \right) \left(1- \frac{\Delta s}{ L}  \right) \right]^2}}
\end{eqnarray}

\begin{eqnarray}\label{gamma M}
\gamma_M = \frac{1}{\sqrt{1-\frac{1}{c^2} \left[ \left( \frac{\Delta s}{ \Delta t} \right) \left(1- \frac{m}{ M} \right) \right]^2}}
\end{eqnarray}

\begin{eqnarray}\label{gamma TLM}
 \gamma_{TLM} = \frac{1}{\sqrt{1-\frac{1}{c^2} \left[ \left( \frac{\Delta s}{\Delta t} \right) \left( 1 - \frac{\Delta t}{T} \right) \left( 1 - \frac{\Delta s}{L} \right) \left( 1 - \frac{m}{M} \right) \right]^2}}
\end{eqnarray}

\noindent The modified special relativistic factors $\gamma_T$, $\gamma_L$, $\gamma_M$ and $v_{TLM}$ may be contrasted with the usual $\gamma_{STR}$ of standard Special Theory of Relativity (STR).

In equation [\ref{Velocity definition T}], the term $\Delta t /T$ can be replaced by any general function satisfying the equations [\ref{Velocity Newtonian limit}] and [\ref{Recurrence constraint}]. It would be interesting to find criteria which would lead to specification of such a function. Similar considerations apply in equations [\ref{Velocity definition L}] and [\ref{Velocity definition M}] and for $\gamma_{TLM}$ of equation [\ref{gamma TLM}].


\begin{thebibliography}{unsrt}

\bibitem{Segal 1976}  Segal I.E. :  \textit{Mathematical   Cosmology   and
Extra-galactic Astronomy}, Cambridge University Press (1976) .

\bibitem{Snyder 1947} Snyder, H.S.:  \textit{Phys. Rev.}, \textbf{72}, (1947), 38 .

\bibitem{Rey 2006} Rey, S. and Hikida, Y.: \textit{Emergent AdS3 and BTZ Black Hole from Weakly Interacting Hot 2d CFT}, JHEP, 0607:023,2006, arXiv:hep-th/0604102v3 .

\bibitem{Kleeban 2004} Kleban, M., Porrati, M. and Rabadan, R.: \textit{Poincare Recurrences and Topological Diversity}, JHEP0410:030,2004, arXiv:hep-th/0407192v2 .

\bibitem{Kim 1999} Kim, S. P. : \textit{Quantum Field Theory in a Topology Changing Universe}, 	Class.Quant.Grav. 16 (1999) 3987-3997, 	arXiv:hep-th/9902077v2 .

\bibitem{Smolin 1995} Smolin, L. and Soo, C. : \textit{The Chern-Simons Invariant as the Natural Time Variable for Classical and Quantum Cosmology}, Nucl.Phys. B449 (1995) 289-316, 	arXiv:gr-qc/9405015v2 .


\bibitem{Boblest 2010} Boblest, S.,  M\"{ü}ller, T., and Wunner, G. : \textit{Twin Paradox in de Sitter Spacetime}, arXiv:1009.3427 . 

\bibitem{Barrow 2001} Barrow, J. D., and Levin, J : \textit{The twin paradox in compact spaces}, Phys.Rev. A63 (2001) 044104, arXiv:gr-qc/0101014 .


\bibitem{Modgil 2009} Modgil, M.S. \textit{Loschmidt's paradox, entropy and the topology of spacetime}, arXiv:0907.3165

\end{thebibliography}
\end{document}